\documentclass[aps,twocolumn,showpacs,floatfix]{revtex4}
\usepackage{graphicx}
\usepackage{dcolumn}
\usepackage{bm}
\usepackage{epsf}
\usepackage{subfigure}
\usepackage{epstopdf}
\usepackage{amsmath}
\usepackage{amssymb}
\usepackage[cspex,bbgreekl]{mathbbol}
\usepackage{color}

\begin{document}

\title{Stochastic thermodynamics of hidden pumps}

\author{Massimiliano Esposito}
\address{Complex Systems and Statistical Mechanics, University of Luxembourg, L-1511 Luxembourg, Luxembourg} 
\author{Juan MR Parrondo}
\address{Departamento de Fisica At\'omica, Molecular y Nuclear and  GISC, Universidad Complutense Madrid, 28040 Madrid, Spain}

\date{\today}

\begin{abstract}
We show that a reversible pumping mechanism operating between two states of a kinetic network can give rise to Poisson transitions between these two states.
An external observer, for whom the pumping mechanism is not accessible, will observe a Markov chain satisfying local detailed balance with an emerging effective force induced by the hidden pump. 
Due to the reversibility of the pump, the actual entropy production turns out to be lower than the coarse grained entropy production estimated from the flows and affinities of the resulting Markov chain. 
Moreover, in presence of a large time scale separation between the fast pumping dynamics and the slow network dynamics, a finite current with zero dissipation may be produced. 
We make use of these general results to build a synthetase-like kinetic scheme able to reversibly produce high free energy molecules at a finite rate and a rotatory motor achieving 100\% efficiency at finite speed.
\end{abstract}


\maketitle

\section{Introduction}


Kinetic networks are a major tool to model physical and chemical systems. They consist of rate equations describing the evolution of the occupancy probability of a given state in the network.  A key quantity to assess the performance of a kinetic process is the entropy production. The efficiency of chemical motors or of biochemical processes such as metabolic cycles, replication, transcription, or proofreading, typically achieves its maximum value when the entropy production is minimal.
Hill, in his classic work \cite{Hill77} on the transduction of free energy in chemical reactions, provided the basic tools to calculate the entropy production of processes modeled by kinetic networks. Closely related results were also found by Schnakenberg \cite{Schnakenberg}.
They showed that the entropy production in a network consists of a sum of positive edge contributions, each expressed as the product of a probability flux across the edge times the edge affinity (or thermodynamic force).   
Their theoretical framework constitutes the basis of stochastic thermodynamics \cite{Sekimoto10, Seifert12Rev, QianPR12a, EspVDBRev2014} which has become central for the study of molecular machines \cite{ParrondoRev02, LacosteMallick07, Seifert11EPJE, Gaspard07JTB}.    

In many applications the observer has only a partial access to the kinetic network. Some states are hidden and the resulting description of the system becomes ``coarse grained''. The thermodynamic implications of coarse graining is an active field of research \cite{JarzynskiRahav07, ParrondoVDBPRE08, VulpianiSTAT10, EspositoCuetaraGaspard11, EspositoBulnessGasp13, EspoDiana14, SeifertBlickleBechingerPRL12, SeifertSpeckJCP13, AltanerVollmerPRL12, ParrondoHoroSag12, CelaniBoJSTAT14}.
The coarse grained entropy production is typically lower than the actual one since it misses the positive contribution of the hidden edges. This result is true for autonomous systems if the hidden variables are even under time reversal. It can be proved in various ways \cite{VandenBroeck07, Esposito12} and remains true even when the coarse grained description is no longer Markovian \cite{ParrondoRoldanPRL10, ParrondoEdgarPRE12, LacosteJCP13}. Indications that odd hidden variables do not satisfy this result were analyzed in the context of Langevin equations where velocities were coarse grained \cite{CelaniPRL12, NakayamaPRE13}.

In this paper we prove that for systems driven by an external time-dependent force, the entropy production at a coarse grained level of description may overestimate the actual entropy production. The driving plays the analogous role of an odd hidden variable when it is not invariant under time reversal. 
We exploit this result to build hyper-efficient pumps. To do so we consider systems with hidden states driven by an external cyclic time-dependent force generating currents between the apparent states. The driving is based on adiabatic pumping previously introduced in the literature \cite{ParrondoPRE98, Astumian01, Astumian03PRL, AstumianPNAS07, Sinitsyn07, HorowitzJarzynskiPRL08} but gives rise in our case to a Markovian dynamics at the coarse grained level.
We show that our reversible pumps can be used to generate currents against a bias with zero entropy production.

\section{Thermodynamics of kinetic networks}

We first review the main results concerning energetics and entropy production in kinetic networks. Consider a system interacting with a reservoir at temperature $T$ and with states $i$ that have energies $E_i$ and are occupied with probabilities $p_i$. The system dynamics is described by a Markovian master equation 
\begin{equation}
\dot{p}_i= \sum_{j} J_{ij} =\sum_j \left[ w_{ij}p_j- w_{ji} p_{i}\right],
\end{equation}
where $J_{ij}$ is the net probability current from site $j$ to site $i$. 
The rates $w_{ij}$ describing the reservoir induced transitions from $j$ to $i$ satisfy local detailed balance
\begin{equation}\label{db}
\ln \frac{w_{ij}}{w_{ji}}=-\beta (E_i-E_{j}-F_{ij}),
\end{equation}
where $\beta^{-1}=k T$ and $F_{ij}$ is a non-conservative thermodynamic force pointing from $j$ to $i$ which could be induced for instance by a non-equilibrium chemical reservoir or a non-conservative mechanical force. The state energies $E_i$ may change in time, $\dot{E}_i \neq 0$, due to the action of an external agent. For this generic scenario an unambiguous formulation of non-equilibrium thermodynamics ensues \cite{Seifert12Rev,EspVDBRev2014,Esposito12}. 

The first law of thermodynamics reads
\begin{equation}
dE = \delta W_{\rm dr}+\delta W_{\rm nc}+\delta Q  \label{firstlaw}
\end{equation}
and expresses the fact that the average energy of the system, $E=\sum_i E_ip_i$, changes due to three mechanisms: 
the driving work corresponding to the energy transferred to the system by the external agent
\begin{eqnarray}
\delta W_{\rm dr}=\sum_i p_i dE_i = \sum_i p_i \frac{d E_i}{dt}  dt ,\label{drw}
\end{eqnarray}
the non-conservative work corresponding to the energy transferred to the system by the non-conservative forces
\begin{eqnarray}
\delta W_{\rm nc}=\sum_{i<j} J_{ij} F_{ij} dt \label{ncw},
\end{eqnarray}
and the heat corresponding to the energy transferred from the reservoir to the system which, using (\ref{db}), can be written as
\begin{equation}
\delta Q = \sum_i E_i dp_i - \delta W_{\rm nc} =-kT \sum_{i<j} J_{ij}\ln\frac{w_{ij}}{w_{ji}}\, dt . \label{heat}
\end{equation}

The second law expresses the fact that the change in the total entropy or entropy production (i.e. the sum of the change in the system Shannon entropy $S = -k \sum_i p_i\ln p_i$ plus the change in the entropy of the reservoir $-\delta Q/T$) is always nonnegative $\delta S_{\rm tot} \geq 0$:
\begin{equation}\label{entropyprod}
\delta S_{\rm tot}=dS-\frac{\delta Q}{T}=\sum_{i<j} \delta S^{ij}_{\rm tot} ,
\end{equation}
where
\begin{equation}\label{sij}
\delta S^{ij}_{\rm tot}=k J_{ij} \ln \frac{w_{ij}p_j}{w_{ji}p_i}\,dt \geq 0
\end{equation}
is the nonnegative edge entropy production expressed as a flux times a force \cite{Hill77,Schnakenberg}.
The total entropy production may also be rewritten as
\begin{equation}\label{dissipativework}
T\delta S_{\rm tot}=\delta W_{\rm dr}+\delta W_{\rm nc}-d {\cal F} = \delta W_{\rm nc}-\sum_{i<j} \delta {\cal F}_{ij},
\end{equation}
where ${\cal F}=E-TS=\sum_ip_i(E_i+kT\ln p_i)$ is the nonequilibrium free energy of the system whose change can in turn be split as 
\begin{equation}\label{FreeEnCh}
d{\cal F} = \delta W_{\rm dr} + \sum_{i<j} \delta {\cal F}_{ij},
\end{equation}
with
\begin{equation}\label{fij}
\delta {\cal F}_{ij}\equiv J_{ij}\,\left[E_i-E_j+kT\ln\frac{p_i}{p_j}\right] \, dt.
\end{equation}
Note that, in the absence of non-conservative forces, $F_{ij}=0$, 
\begin{equation}\label{EPnoNCF}
T\delta {S}^{ij}_{\rm tot}=-\delta {\cal F}_{ij} \ \ \;, \ \ T\delta S_{\rm tot}=-\sum_{i<j} \delta {\cal F}_{ij}.
\end{equation}

\section{Inducing Poissonian transitions by reversible pumping}

We now introduce a generic reversible pumping  mechanism that transfers probability between two states in a way that is indistinguishable from the Poissonian transitions of a Markovian dynamics. Poissonian transitions are characterized by the following properties: {\em i)} the probability transferred during a small time interval $\tau$ is $w \tau$, where $w$ is the rate of the transition; and  {\em ii)} the occurrence of a transition in a given time interval is independent from the transitions that occurred in the past. When considering a pump induced by a periodic driving of very small period $\tau$ (this condition will be made more precise below) and transferring a probability $w \tau$ during each cycle, then the pump will mimic Poissonian transitions, since the transitions that occurred in a given cycle are independent of those occurring in the other ones.

\begin{figure}[h]
\center{\rotatebox{0}{\scalebox{0.25}{\includegraphics{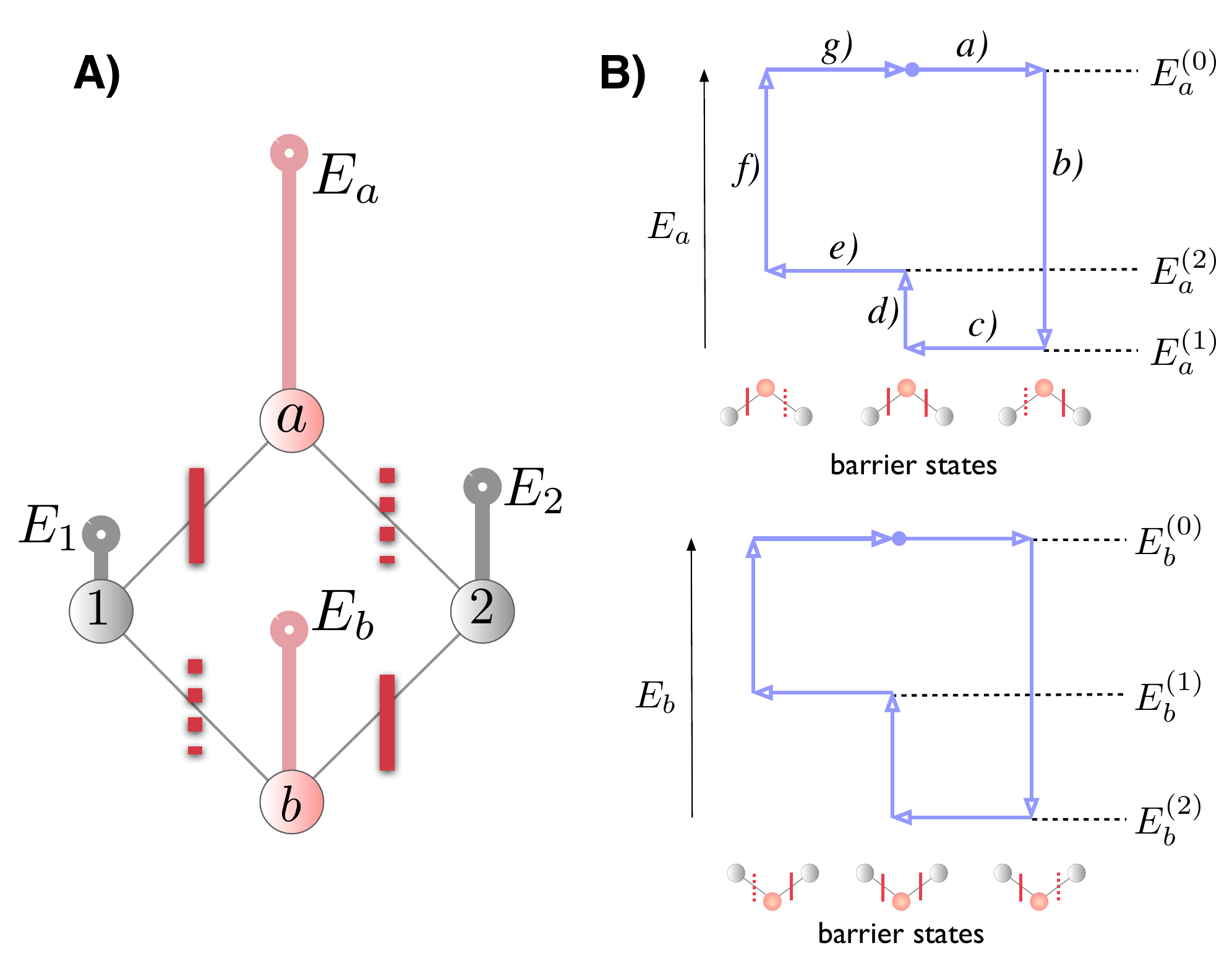}}}} 
\caption{{\bf Schematic representation of the reversible pump}. A) The pump between network states 1 and 2 consists of two intermediate states $a$ and $b$ with respective energies $E_a$ and $E_b$, which are modified by an external agent in a cyclic way. The agent can also open (solid vertical bars) and close (dashed vertical bars) the barriers connecting the network states $1$ and $2$ with the pump states $a$ and $b$. B) Protocol followed by pump $a$ (upper figure) and pump $b$ (lower figure). We have labeled the 7 steps of the $a$ protocol, according to the description in the main text. Notice that the superscript of the energies indicates the network state $i$ which is in contact with the pump state $a,b$ when $E_{a,b}=E^{(i)}_{a,b}$. The pump is reversible if the changes in the energies $E_{a,b}$ are carried out quasistatically with respect to the time scale of the transitions between network states and pump states and if $E^{(2)}_a$ and $E^{(1)}_b$ are appropriately chosen (see (\ref{e2eq}) and (\ref{e2eq2})). The opening and closing of the barriers can be done instantaneously without compromising the reversibility of the process.}
\label{Fig1}
\end{figure}

To be precise, consider the setup depicted in Fig.~\ref{Fig1} A. The system is made of observable states $1,2,3,\dots$ ($3,4,\dots$ not shown in the figure) and two hidden states $a,b$ connecting states $1,2$. The transition rates satisfy local detailed balance (\ref{db}). The transitions between $a,b$ and $1,2$ can be turned on and off by an external agent without any expenditure of work (this can be achieved for instance for Arrhenius rates $w_{ij}=\Gamma_{ij} e^{\beta E_j}$ by instantaneously raising and lowering the energy barriers $\Gamma_{ij}=\Gamma_{ji}$) and do not involve any non-conservative forces $F_{ij}=0$. The external agent also controls the two energies $E_a$ and $E_b$. The operations performed by the external agent are cyclic and of period $\tau$ chosen to be much shorter than any time scale between the observable states, i.e. $\tau w_{ij} \ll 1$ for $i,j=1,2,3,\dots$. 

We first describe the process along path $1-a-2$ where the energy $E_a$ and the barriers between $a$ and $1$ and between $a$ and $2$ are modulated. The protocol starts with the two barriers closed and an energy $E_a=E_a^{(0)}\gg E_1,E_2$, and proceeds as follows (see Fig.~\ref{Fig1} B): 
a) the barrier $1-a$ is opened; 
b) the energy $E_a$ is quasistatically lowered to $E_a^{(1)}$; 
c) the barrier $1-a$ is closed; 
d) the energy $E_a$ is changed to $E_a^{(2)}$; 
e) the barrier $a-2$ is opened; 
f) the energy $E_a$ is quasistatically restored to its initial value $E_a^{(0)}$; 
g) the barrier $a-2$ is closed to complete the cycle. 
We note that, while the barriers can be opened or closed instantaneously, the changes in $E_a$ are carried out quasistatically to minimize the entropy production, except for step d) where state $a$ is not connected with states $1$ and $2$ and $E_a$ can be changed arbitrarily fast without compromising the reversibility of the cycle. 

The cycle is engineered in such a way that site $a$ is practically empty at the beginning and at the end of the cycle. 
Therefore, initially $p_a=0$ and the probability to be on state $1,2$ is denoted by $p_1,p_2$. 
During step a) an irreversible probability leak occurs from $p_a=0$ and $p_1$ to
\begin{equation}
p'_a=\frac{p_1\,e^{-\beta (E_a^{(0)}-E_1)}}{1+e^{-\beta (E_a^{(0)}-E_1)}};\quad
p'_1=\frac{p_1}{1+e^{-\beta (E_a^{(0)}-E_1)}}.
\end{equation}
We assume $\beta(E_a^{(0)}-E_1)\gg 1$ and neglect this leak: $p'_1\simeq p_1$ and $p'_a\simeq 0$ (see the discussion below on the different scales of energy and time in our model). 
During step b), a quasistatic reversible transfer of probability from state $1$ to state $a$ is performed. Because the two states are in equilibrium with respect to each other, the respective occupation probabilities after step b) are
\begin{equation}
p''_a=\frac{p_1\,e^{-\beta (E_a^{(1)}-E_1)}}{1+e^{-\beta (E_a^{(1)}-E_1)}};\quad
p''_1=\frac{p_1}{1+e^{-\beta (E_a^{(1)}-E_1)}}.
\end{equation}
Since, after step b), the barrier $1-a$ remains closed for the rest of the cycle, $p''_a$ is the probability that will be transferred from site 1 to site 2 after the cycle is completed. For the pump to mimic Poisson transitions, this probability must be of the order of the duration of the cycle $\tau$. Therefore, we impose the following scaling relationship between $\tau$ and the energy $E_a^{(1)}$
\begin{equation}\label{scaling}
e^{-\beta (E_a^{(1)}-E_1)}=w_{21} \tau,
\end{equation}
where $w_{21}$ is a finite rate which we will soon prove to be the effective rate of transitions from state 1 to state 2. 
For the transitions to be Poissonian we have to further impose that $w_{21} \tau \ll 1$, which amounts to impose $\beta (E_a^{(1)}-E_1) \gg 1$. In the following, we approximate all the expressions up to first order in $\tau$, since this is the approximation that yields a Markovian dynamics ruled by an effective master equation, once the pump is coarse grained.
Notice also that the initial energy $E_a^{(0)}$ should be even bigger than $E_a^{(1)}$ since it must lead to $e^{-\beta (E_a^{(0)}-E_1)} \ll w_{21}\tau$ in order to justify neglecting the irreversible leak of step a). 
Using the scaling (\ref{scaling}), the transferred probability from $1$ to $a$ to first order in $\tau$ reads
\begin{equation}
p''_a \simeq p_1e^{-\beta (E_a^{(1)}-E_1)}=p_1 w_{21} \tau.
\end{equation}
We now impose the following relation on $E_a^{(2)}$
\begin{equation}\label{e2eq}
\frac{p''_a}{p_2}=e^{-\beta (E_a^{(2)}-E_2)}
\end{equation}
which is equivalent to
\begin{equation}\label{e2eq2}
E_a^{(2)}-E_a^{(1)}=-kT\ln\frac{p_1}{p_2}+E_2-E_1.
\end{equation}
As a result, the probabilities $p''_a$ and $p_2$ are in equilibrium when the barrier $a-2$ is opened in step e). Hence, the entropy production is  zero along this step as well as along step f) and g). Beside the initial probability leak in step a), which can be made arbitrary small, the remaining steps are reversible. 
We also note that in all previous expressions, the shifts in $p_1$ and $p_2$, due to transitions with other observable states $3,4,\dots$ or due to the dynamics along path $1-b-2$, are of order $\tau$. They will therefore only affect terms of second order in $\tau$ and will not prevent the entropy production of the process to vanish up to first order in $\tau$.

We now turn to evaluating the work performed by the external agent on the system when changing the energy $E_a$ along steps b), d), and f). The remaining steps a), c), and e) involve no work since only the barriers are changed. As every step involving work is quasistatic and reversible, the driving work can be calculated as a difference of equilibrium free energy.
We find
\begin{eqnarray}
&&\hspace{-0.7cm} W_b =-p_1kT\left[\ln \left(e^{-\beta E_a^{(1)}}+e^{-\beta E_1}\right)-\ln \left(e^{-\beta E_1}\right)\right] \simeq -kTp_a'' \nonumber \\
&&\hspace{-0.7cm} W_d = p''_a\left[E_a^{(2)}-E_a^{(1)}\right] \\
&&\hspace{-0.7cm} W_f =-p_2kT\left[\ln \left(e^{-\beta E_2}\right)-\ln \left(e^{-\beta E_a^{(2)}}+e^{-\beta E_2}\right)\right] \simeq kTp_a'' \nonumber . 
\end{eqnarray}
The overall work along path $1-a-2$ can thus be written as
\begin{eqnarray}
W^{(a)} = p''_a\left[E_a^{(2)}-E_a^{(1)}\right] = p''_a\left[E_2-E_1-kT\ln\frac{p_1}{p_2}\right] .\label{wa}
\end{eqnarray}
The r.h.s.of this equation is the change of free energy in the system due to the probability $p_a''$ transferred by the pump, confirming that the entropy production due to the pumping mechanism vanishes. 

We now turn to the process affecting path $1-b-2$. The energy $E_b$ and the barriers between $b$ and $1$ and between $b$ and $2$ are changed in a similar way as along path $1-a-2$ (see Fig.~\ref{Fig1} C).
The analysis for this part of the protocol is analogous to that of $1-a-2$, and the resulting expressions are obtained by just swapping $a$ and $b$ as well as $1$ and $2$.

By combining the results obtained along the two paths, $1-a-b$ and $2-b-1$, we find the first important result of this paper, namely that the effective rates $w_{21}$ and $w_{12}$ satisfy a local detailed balance relation (\ref{db}) 
\begin{equation}\label{ModLDB}
\frac{w_{21}}{w_{12}}=e^{-\beta(E_2-E_1-F^{\rm eff}_{21})},
\end{equation}
which, contrary to the original rates, now contains an effective non-conservative force
\begin{equation}
F^{\rm eff}_{21} \equiv E_b^{(2)}-E_a^{(1)} \label{effectiveforce}
\end{equation}
pointing from $1$ to $2$. 
Furthermore the total work performed by the pump during a cycle is given by
\begin{eqnarray}
&&\hspace{-0.5cm}\delta W_{\rm dr}=W^{(a)}+W^{(b)} = p_a''\left[E_a^{(2)}-E_a^{(1)}\right]+p_b''\left[E_b^{(1)}-E_b^{(2)}\right] \nonumber \\
&&\hspace{0.3cm} = (p_a''-p_b'')\left[-kT\ln\frac{p_1}{p_2}+E_2-E_1\right] \nonumber \\
&&\hspace{0.3cm} = J_{21}\tau\,\left[-kT\ln\frac{p_1}{p_2}+E_2-E_1\right] = \delta{\cal F}_{21}=\delta{\cal F}_{12} ,\label{wdrtot}
\end{eqnarray}
where we used (\ref{fij}) with $d t=\tau$ for the last equality.

\section{Coarse grained versus real entropy production}

We now turn to the comparison between the real entropy production of the full network which includes the pumping states and the coarse grained entropy production obtained by just considering the dynamics on the observable states. For simplicity, we assume no non-conservative force besides the effective force $F^{\rm eff}_{21}$ emerging at the coarse grained level. Examples with non-conservative forces will be provided in the applications. We consider pumping cycles of duration $\tau$ much smaller than the characteristic time of the dynamics of the coarse grained network.

At the coarse grained level of description, the observed states are not driven and the only non-conservative force is the effective one induced by the pump. The total work in a cycle is therefore $\delta W_{\rm nc} = J_{12} F^{\rm eff}_{12} \tau$ and, using (\ref{dissipativework}), the entropy production per cycle reads
\begin{equation}\label{EPcycleCG}
T\delta S_{\rm tot}^{\rm (cg)}=J_{12}F^{\rm eff}_{12} \,\tau-\sum_{i<j}\delta{\cal F}_{ij} \geq 0,
\end{equation}
where the sum runs over the observable states $i,j=1,2,3,\dots$ and $\delta {\cal F}_{ij}$ is given by Eq.~(\ref{fij}) with $d t=\tau$. 

On the other hand, in the full network all forces are conservative. Using (\ref{dissipativework}) and (\ref{wdrtot}), the true entropy production is given by
\begin{equation}\label{EPcycleInt}
T\delta S_{\rm tot} = \delta W_{\rm dr}-d{\cal F} = \delta{\cal F}_{12}-d{\cal F} \geq 0.
\end{equation}
When calculating the differential $d{\cal F}$ over a cycle of the pump operation, the contributions to ${\cal F}$ from the hidden states $a,b$ vanish since they are empty at the beginning and at the end of the cycle. Since the remaining states do not depend on the external agent, one finds
\begin{equation}\label{dFcycle}
d{\cal F}=d\left [\sum_{i} (E_{i}+kT\ln p_{i})p_{i}\right]=\sum_{i<j}\delta{\cal F}_{ij}.
\end{equation}
Inserting (\ref{dFcycle}) in (\ref{EPcycleInt}), we get that
\begin{eqnarray}\label{si0}
T\delta S_{\rm tot} = -\sum_{i<j \neq 2}\delta{\cal F}_{ij} \geq 0.
\end{eqnarray}
Comparing this result with (\ref{EPnoNCF}), we observe that the link $1-2$ does not contribute to the entropy production, confirming the reversibility of the pumping mechanism. 

Our second important result is that the true entropy production overestimates the coarse grained one: 
\begin{eqnarray}\label{KeyIneq}
\delta S_{\rm tot}^{\rm (cg)} \geq \delta S_{\rm tot}.
\end{eqnarray}
This result follows from comparing (\ref{EPcycleCG}) with (\ref{si0}) using the inequality 
\begin{eqnarray}
k T J_{12} \ln \frac{w_{12}p_2}{w_{21}p_1} dt = J_{12} F_{12}^{\rm eff} dt - \delta{\cal F}_{12} \geq 0.
\end{eqnarray}

Of special interest is the entropy production rate when the system reaches a stationary state. In this case, $d{\cal F}=0$ in (\ref{dissipativework}), and the entropy production in the coarse grained network is given by the non-conservative work, whereas the real entropy production is proportional to the driving work (\ref{wdrtot}). The respective entropy production rates are:
\begin{eqnarray}
T\dot S_{\rm tot}^{\rm (cg)} &\equiv& T \frac{\delta S_{\rm tot}^{\rm (cg)}}{\tau}= J_{21}\,F_{21}^{\rm eff} \label{sicg} \\
T\dot S_{\rm tot} &\equiv& T \frac{\delta S_{\rm tot}}{\tau} 
= \frac{\delta{\cal F}_{12}}{\tau} = \frac{\delta W_{\rm dr}}{\tau} \nonumber \\
&=& J_{21}\left[E_2-E_1+kT\ln\frac{p_2}{p_1}\right] \label{si} .
\end{eqnarray}
Note that $\dot S_{\rm tot}$ may vanish even for a finite current $J_{21}$ (an example is provided below). 

The driving protocol that we have introduced to pump reversibly between a pair of observable states can be designed for any system with preassigned effective rates $w_{21},w_{12}$ and operating in the stationary regime.
Indeed, the choice of $w_{21}$ and $w_{12}$ determines the effective force $F^{\rm eff}_{21}$ via (\ref{ModLDB}), and along with the rest of the Markov chain, also determines the stationary values of $p_1$ and $p_2$. From these stationary values we set $E_a^{(2)}$ and $E_b^{(1)}$ using Eq.~(\ref{e2eq2}) and we set $E_a^{(1)}$ and $E_b^{(2)}$ using (\ref{scaling}). 
It should be clear that our procedure can be easily generalized to systems with pumps located between several pairs of observable states and/or to systems with non-conservative forces besides the ones induced by the pumps. Some examples of this are provided below.

\section{Applications}\label{applications}

\subsection{Pump embedded in a conservative network}

\begin{figure}[h]
\center{\rotatebox{0}{\scalebox{0.25}{\includegraphics{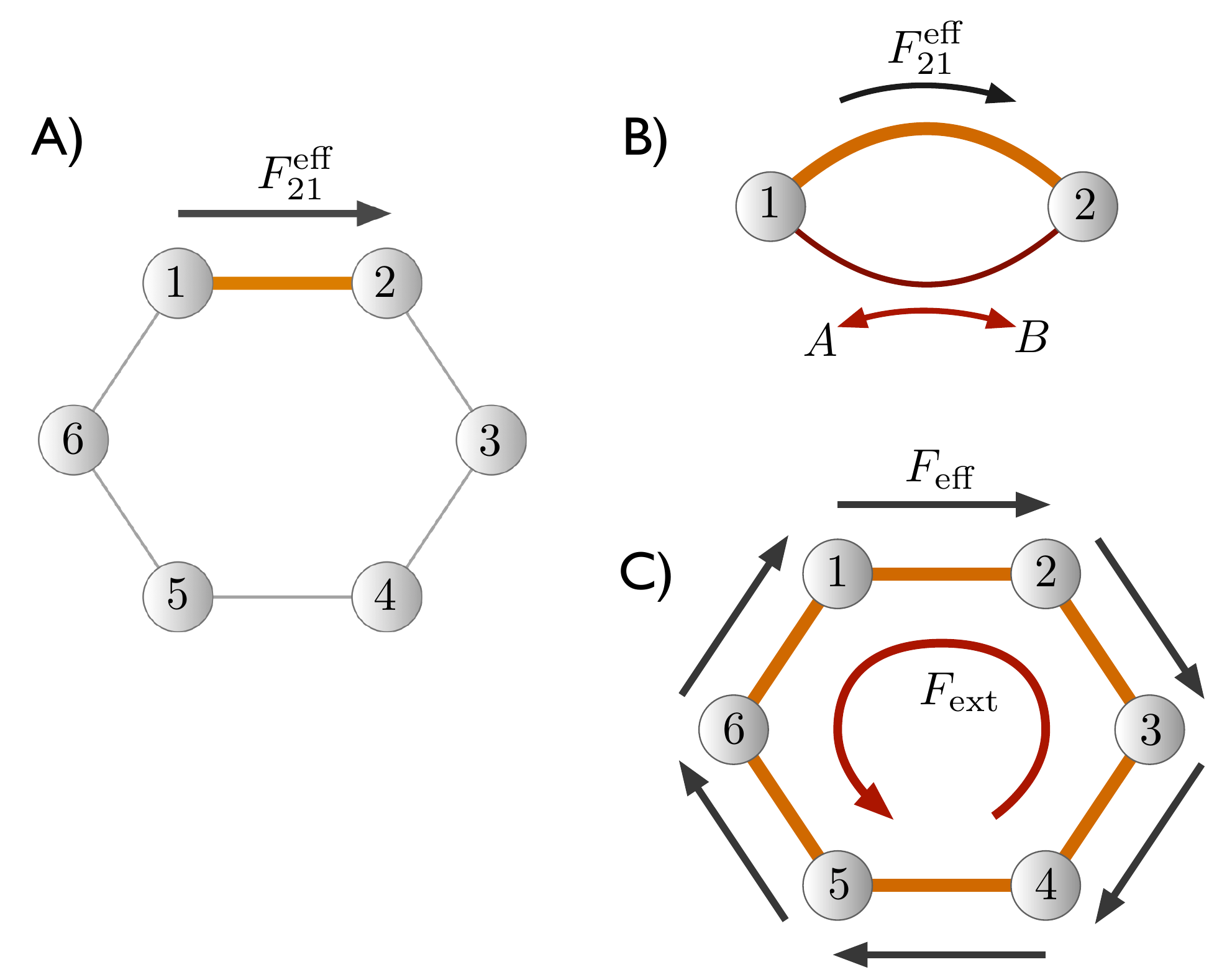}}}}
\caption{{\bf Three examples}. The thick colored link indicates the presence of a hidden pump inducing a force $F^{\rm eff}$ in the direction of the black arrow. The examples are: A) A ring with a pump connecting two network states, 1 and 2. B) A kinetic network that produces high free energy molecules $A$ from low free energy molecules $B$ ($\Delta\mu=\mu_A-\mu_B>0$). C) A ring with pumps at every link, working against a uniform force $F_{\rm ext}$.}
\label{Fig2}
\end{figure}

As a first example, we consider a system of $N$ states $i=1,2,\dots,N$ connected as a ring (see Fig.~\ref{Fig2} A). The states energies are all zero $E_{i}=0$, no non-conservative forces act on the system, and a hidden pump is inserted between states $1$ and $2$. Local detailed balance implies equal rates $w_{i\,i\pm 1}=w$ for all transitions except $w_{21}$ and $w_{12}$ which satisfy $\ln (w_{21}/w_{12})=\beta F^{\rm eff}_{21}$, where $F^{\rm eff}_{21}$ is the effective force induced by the pump.
The stationary state of the system is 
\begin{eqnarray}
p_1 &=& \frac{[(N-1)w_{12}+w]J}{(w_{21}-w_{12})w} \nonumber \\
p_{n} &=& p_1+(N+1-n)\frac{J}{w}, \ \ n=2,\dots,N,
\end{eqnarray}
where the clockwise stationary current $J$ is given by
\begin{equation}
J \equiv J_{21}=\frac{2w[w_{21}-w_{12}]}{2Nw+N(N-1)(w_{21}+w_{12})}.
\end{equation}
The real entropy production rate is proportional to the driving work performed by the pump and reads (see (\ref{si}))
\begin{equation}
T \dot S_{\rm tot} = kT J \ln\frac{p_2}{p_1} = kT J \ln \left[\frac{(N-1)w_{21}+w}{(N-1)w_{12}+w}\right],
\end{equation}
whereas the coarse grained entropy production rate is given by (see (\ref{sicg}))
\begin{equation}
T \dot S_{\rm tot}^{\rm (cg)} = kT J \ln \left( \frac{w_{21}}{w_{12}}\right) = J F^{\rm eff}_{21}.
\end{equation}

If the network transitions are much slower than the effective rates from 1 to 2, i.e. if $w\ll w_{12},w_{21}$, then the coarse grained entropy coincides with the real one $\dot S_{\rm tot} \simeq \dot S_{\rm tot}^{\rm (cg)}$. The same is true if $N \to \infty$.
On the other hand if $w\gg w_{12},w_{21}$, the real entropy production vanishes despite the fact that the network exhibits a finite dissipative current $J$ giving rise to an apparent entropy production $\dot S_{\rm tot}^{\rm (cg)} = J F^{\rm eff}_{21}$. 

These results generalize to any conservative network obeying detailed balance and containing a hidden pump in the edge $1-2$. If the rates along the normal edges are much larger than the effective rates along the pumping edge ($w_{12}$ and $w_{21}$), the whole chain will be at equilibrium with respect to the energy landscape $E_i$, $i=1,2,\dots$. Then, according to (\ref{si}), $\dot S_{\rm tot}=0$, whereas a finite current $J_{21}=w_{21}p_1-w_{12}p_2$ gives rise an apparent entropy production $\dot S_{\rm tot}^{\rm (cg)}=J_{21} F^{\rm eff}_{21}$. This remarkable result is not in contradiction with any fundamental law of thermodynamics. The dissipationless finite current arises from the large separation of time scales: the current is finite over the time scales $1/w_{ij}$ of the dynamics on the observable states but is induced quasistatically over the internal time scale of the pump. A similar phenomenon has been previously discussed in the context of adiabatic pumps \cite{ParrondoPRE98, Astumian03PRL, AstumianPNAS07, Sinitsyn07, HorowitzJarzynskiPRL08}.

\subsection{A highly efficient synthetase}

We now consider an enzyme switching between two conformational states $1$ and $2$ with the same energy $E_1=E_2=0$. The enzyme jumps due to two different mechanisms with respective rates $w^{\rm pump}_{ij}$ and $w^{\rm reac}_{ij}$. The first is induced by a hidden pump generating an apparent effective force $F^{\rm eff}_{21}= kT \ln (w^{\rm pump}_{21}/w^{\rm pump}_{12})$ from $1$ to $2$, and the second is induced by a chemical reaction $1+A\leftrightarrow 2+B$ such that $\Delta \mu=\mu_A-\mu_B=kT \ln (w^{\rm reac}_{21}/w^{\rm reac}_{12})>0$ (see Fig.~\ref{Fig2} B). When operating alone, both mechanisms favor their respective transition towards state $2$. However, when operating simultaneously with $F^{\rm eff}_{21}>\Delta \mu$, the pump can revert the spontaneous direction of the chemical reaction and thus generate high free energy molecules $A$ at a rate given by the (clockwise) stationary current
\begin{equation}
J \equiv w^{\rm pump}_{21}p_1 -w^{\rm pump}_{12}p_2=w^{\rm reac}_{12}p_2-w^{\rm reac}_{21}p_1,
\end{equation}
where the stationary probabilities read
\begin{eqnarray}
p_1 = 1-p_2 = \frac{w^{\rm pump}_{12}+w^{\rm reac}_{12} } {w^{\rm pump}_{12}+w^{\rm reac}_{12}+w^{\rm pump}_{21}+w^{\rm reac}_{21}} .
\end{eqnarray}
The current may be rewritten as 
\begin{equation}\label{j_example2}
J =\frac{w^{\rm reac}_{12} w^{\rm pump}_{21} \left[1-e^{-\beta(F_{21}^{\rm eff}-\Delta\mu)}\right]}
{w^{\rm pump}_{12}+w^{\rm reac}_{12}+w^{\rm pump}_{21}+w^{\rm reac}_{21}}.
\end{equation}

The coarse grained and the real entropy production are obtained by adding the non-conservative work $\delta W_{\rm nc}=-J\Delta \mu$ to Eqs. (\ref{sicg}) and (\ref{si}) respectively:
\begin{eqnarray}
T\dot S^{\rm (cg)}_{\rm tot} &=& J(F_{21}^{\rm eff}-\Delta\mu) \label{scgex} \\
T\dot S_{\rm tot} &=& \delta W_{\rm nc} + \delta W_{\rm dr} =-J\Delta\mu+JkT\ln\frac{p_2}{p_1} \label{sex} \\
&=& JkT\ln\frac{w^{\rm pump}_{21}e^{-\beta \Delta \mu}+w_{12}^{\rm reac}}{w^{\rm pump}_{21}e^{-\beta F_{21}^{\rm eff}}+w_{12}^{\rm reac}} . \nonumber
\end{eqnarray}
The real entropy production ranges  from reversibility, $\dot S_{\rm tot}=0$, if $w^{\rm pump}_{21}\ll w_{12}^{\rm reac}$, to $\dot S_{\rm tot}=\dot S_{\rm tot}^{\rm (cg)}$ if the pump transfers probability much faster than the reaction, $w^{\rm pump}_{21}\gg w_{12}^{\rm reac}$. In the former case it is therefore possible to produce molecules of $A$ with very high efficiency since the synthetase can work at finite rate with a vanishing entropy production. As mentioned before, this does no contradict the second law of thermodynamics since the current occurs on much slower time scale than the driving. One can even show that, for fixed $F^{\rm eff}_{21}$, the efficiency at maximum power (over $\Delta \mu$) tends to 1 when $w^{\rm pump}_{21}/w^{\rm reac}_{12} \to 0$.

To demonstrate that the reversible behavior can be achieved for a large but reasonable separation of time scales, we numerically solved the master equation of the synthetase using Arrhenius rates for the pumping rates $w_{ij}^{\rm pump}=\Gamma_{ij} e^{\beta E_j}$, where $\Gamma_{ij}=\Gamma_{ji}$. This will also help us to show in detail how to build a reversible pump to be inserted in a given network. Energy units are measured in $kT$ and time units in $1/w^{\rm pump}_{21}$. 
We consider a reaction with a difference in chemical potential between species $A$ and $B$ of $\Delta\mu=0.5$ and with rates ranging between $w^{\rm reac}_{21}=0$ -- $10$ and $w^{\rm reac}_{12}=w^{\rm reac}_{21}e^{-\beta \Delta\mu}\simeq 0$ -- 6.065. 

\begin{figure}[h]
\center{\rotatebox{0}{\scalebox{0.25}{\includegraphics{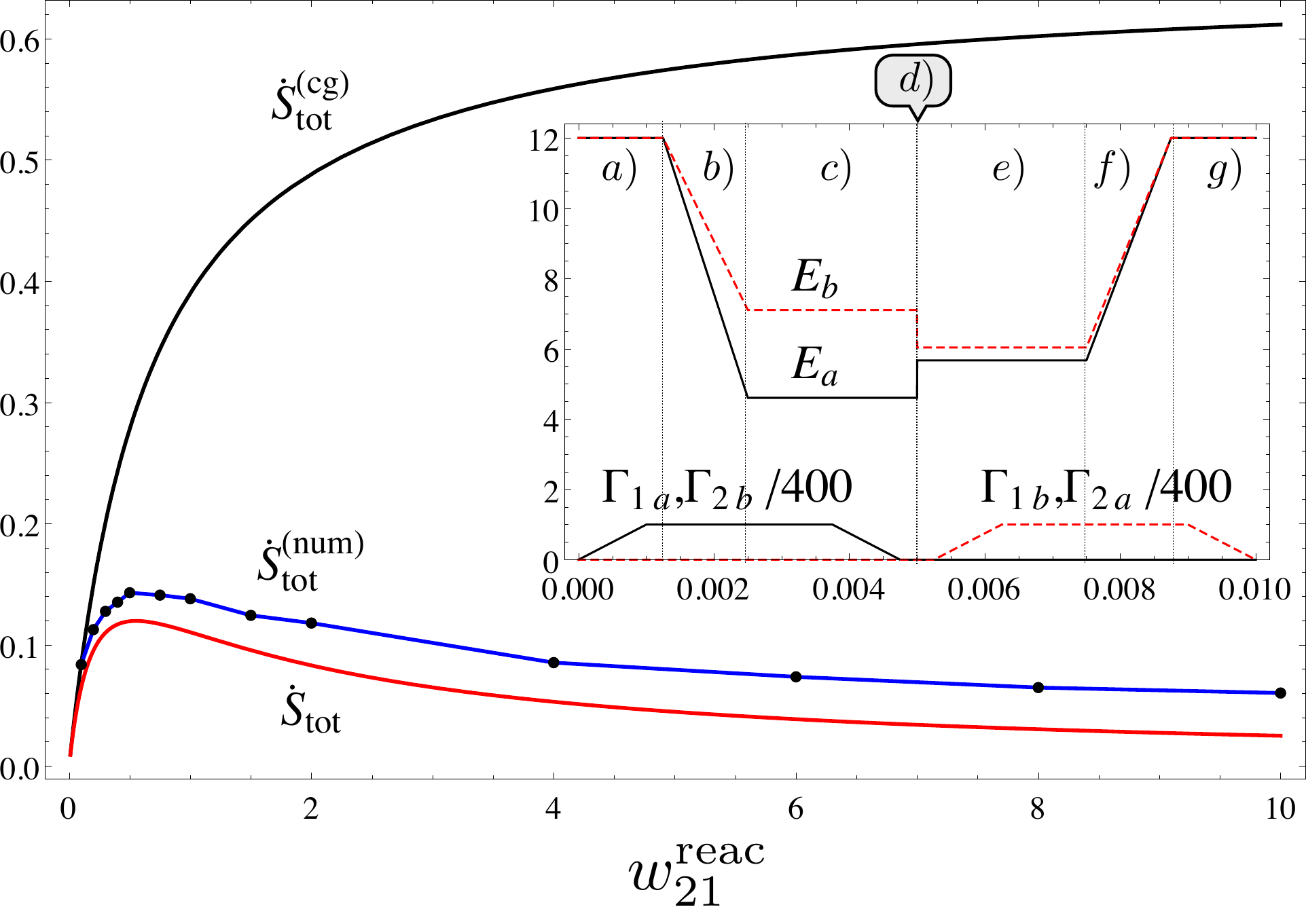}}}}
\caption{
{\bf Numerical solution of the reversible synthetase.} Comparison for the example of Figure \ref{Fig2} B between the coarse-grained entropy production $\dot{S}_{\rm tot}^{(\rm cg)}$ given by (\ref{scgex}), the entropy production of the ideal reversible pump $\dot{S}_{\rm tot}$ given by (\ref{sex}), and the entropy production calculated numerically $\dot{S}_{\rm tot}^{(\rm num)}$ for $\Gamma \simeq 400$ and $\tau=0.01$ in units of energy $kT=1$ and time $1/w^{\rm pump}_{21}=1$. The pump is build to produce high chemical potential $A$ molecules $\Delta \mu=\mu_A-\mu_B=0.5$ and a pump force $F_{21}^{\rm eff}=2.5$. The reaction rates producing $A$ range from $w^{\rm reac}_{21}=0$ -- $10$ and from $w^{\rm reac}_{12}=w^{\rm reac}_{21}e^{-\beta \Delta\mu} \simeq 0$ -- $6.065$. Given these parameters, the protocol followed by the energies of the internal states $E_a, E_b$ is fully determined, as explained in the main text. The protocol for the energies and barriers is depicted in the inset.}
\label{Figsim}
\end{figure}

Our goal is to build a pump exerting a force $F^{\rm eff}_{21}=2.5$ with effective rates $w^{\rm pump}_{21}=1$ and $w^{\rm pump}_{12}=w^{\rm pump}_{21}e^{-\beta F^{\rm eff}_{21}}=e^{-2.5} \simeq 0.082$ so that the system will produce $A$ molecules at a rate $J$ ranging between 0 (for $w^{\rm reac}_{21}=0$) and 0.3 (for $w^{\rm reac}_{21}=10$), as obtained from Eq.~(\ref{j_example2}). 
To do so, we first set the cycle time to $\tau=0.01$, i.e. small enough for the pump to generate Poisson rates at the coarse-grained level. 
According to (\ref{scaling}) and the equivalent equation for pump $1-b-2$, this together with $E_1=E_2=0$ fixes the energy of the hidden states $a$ and $b$ at the end of step b) to $E_a^{(1)}=E_1-kT\ln(w_{21}^{\rm pump}\tau)=-\ln\tau\simeq 4.6$ and $E_b^{(2)}=E_2-kT\ln(w_{12}^{\rm pump}\tau)=F^{\rm eff}_{21}-\ln\tau\simeq 7.1$. 
We then fix the energies after step d) according to (\ref{e2eq2}) to $E_a^{(2)}=E_a^{(1)}-\ln(p_1/p_2)$ and $E_b^{(1)}=E_b^{(2)}-\ln(p_2/p_1)$ which depend on the specific value of $w^{\rm reac}_{21}$. For instance for $w^{\rm reac}_{21}=1$, we get that $E_a^{(2)} \simeq 5.67$ and $E_b^{(1)}=6.04$. 
Finally, we set the time scale of the internal transitions in the pump by fixing the value of the open barriers $\Gamma_{ii'}$ ($i=1,2$ and $i'=a,b$). In our numerical solution we open and close the barriers using linear ramps ranging from $400$ to $0$. The protocol for the energies and barriers is depicted in the inset of Figure \ref{Figsim}. 
The entropy production of the system $\dot{S}_{\rm tot}^{(\rm num)}$ obtained by full numerical integration (black points connected by blue lines) is depicted in Fig.~\ref{Figsim}.
It approaches, but still differs from, the entropy production of the ideal reversible pump $\dot{S}_{\rm tot}$ (red curve) and is clearly below the coarse-grained entropy production $\dot{S}_{\rm tot}^{(\rm cg)}$ (black curve). 
The irreversibility in the pump causing the discrepancy between $\dot{S}_{\rm tot}^{(\rm num)}$ and $\dot{S}_{\rm tot}$ mainly occurs at the end of step b) and the beginning of step f). 

\subsection{A reversible rotatory motor}

Our final example is a $N$-state ring with energies $E_i=0$. Each edge contains a hidden pump generating a force $F^{\rm eff}_{i+1,i}=F_{\rm eff}$ (clockwise) and is subjected to a constant external torque $F_{\rm ext}$ (counterclockwise) operating against the pumps (see Fig.~\ref{Fig2} C). If all the pumps are identical, then the stationary state is uniform $p_i=1/N$ and the (clockwise) current reads
\begin{equation}
J=J_{21}=\frac{w_{21}}{N}\left[1-e^{-\beta(F_{\rm eff}-F_{\rm ext})}\right].
\end{equation}
It is positive for $F_{\rm eff}> F_{\rm ext}$ meaning that the pumps generate a finite speed rotation against the torque. 

As in the previous example, the coarse-grained entropy production can be derived by adding to Eq.~(\ref{sicg}) the non-conservative work performed on the $N$ edges of the motor $\delta W_{\rm nc}=- N J F_{\rm ext}\tau$:
\begin{equation}
T\dot S_{\rm tot}^{\rm (cg)} = N J (F_{\rm eff}-F_{\rm ext}).
\end{equation}
It is a non-negative quantity which only vanishes at zero power $J =0$. 
The calculation of the real entropy production is more subtle since, contrary to what happens for the synthetase, the external force $F_{\rm ext}$ affects the internal transitions of the pumps, $1-a$, $1-b$, $2-a$, $\ldots$. The actual energy of site $i$ is zero because the effect of the torque is borne by the external force. However, the work performed by the driving in the pump between site $i$ and $i+1$ is given by (\ref{wdrtot}) replacing $E_{i+1}-E_i$ by $F_{\rm ext}$ and $p_i=p_{i+1}$. The total driving work obtained by summing over the $N$ pumps is therefore
\begin{equation}
\delta W_{\rm dr} = N J F_{\rm ext} \tau
\end{equation}
and the real entropy production rate in the stationary regime vanishes
\begin{equation}
\dot S_{\rm tot}=\frac{\delta W_{\rm dr}}{\tau}+\frac{\delta W_{\rm nc}}{\tau}=0.
\end{equation}
Remarkably, this motor is able to operate reversibly against any external torque $F_{\rm ext}$. 

\section{Discussion}

We have proposed a reversible time-dependent driving mechanism (called reversible pump) which can be inserted between any two states of a kinetic network. When coarse grained, this pump gives rise to a forward and backward Poissonian rate between the two states. The ratio of these effective rates satisfies a local detailed balance displaying an emergent nonconservative force. 
Remarkably, these pumps can always be engineered in such a way to operate reversibly when inserted in any steady state kinetic network. 

We found that, contrary to common belief, the coarse grained Markovian kinetics generated by our pumps exhibits an entropy production which is always larger than the real one. We exploit this fact to propose several ``hyper efficient" setups which produce finite currents (and thus finite entropy production) at the coarse grained level while the real entropy production vanishes.  

The origin of this surprising phenomenon is that coarse graining the driving affects the symmetry of the system under time reversal. 
Entropy production is a measure of the  probabilistic distinguishability between a process and its time reversal \cite{EspoDiana14,ParrondoRoldanPRL10,Gaspard04b}. 
To define the time-reversed process one must consider the time-reversed driving. But if the information concerning the driving is lost during the coarse graining procedure as is the case here, the time-reversal operation at the coarse grained level does not relate anymore to the real time-reversal operation at the underlying level. 
A similar phenomenon may occur if hidden variables which are odd under time reversal are considered, such as velocity, angular momentum, or magnetic moment \cite{NakayamaPRE13, CelaniPRL12}. In fact, an external driving can be implemented by a large mass with a non zero initial velocity \cite{JarzynskiDeffnerPRX13}. 

Our setup has also an intriguing relation with information engines that use the information gathered in a measurement to extract work, in the spirit of the celebrated Maxwell demon.
In Ref.~\cite{ParrondoHoroSag12}  a driven kinetic scheme that works as a Maxwell demon was introduced. When the demon is coarse grained, the resulting dynamics is Markovian and mimics the dynamics of a chemical motor. The scheme is not able to always work reversibly and it is more restrictive than the one presented here, but the demon is able to run the motor with less entropy production than chemical fuel. In this case, the hidden states are long lifetime states (with respect to the internal time scale of the motor) featuring the strong correlation between the motor and the demon implied by the measurement \cite{ParrondoHoroSag12}. It would be interesting to find whether our general scheme also admits an interpretation in terms of information.

Our pumping mechanism is based on two ingredients, the presence of time asymmetric driving (an ``odd variables'') and a large separation of time scales. These can yield a dramatic enhancement of the performance of a kinetic network. It is an open question whether these two ingredients can be helpful for designing more efficient chemical motors and nanodevices or whether they are already present in protein motors and other biological processes.

\section*{Acknowledgments}

M.E. is supported by the National Research Fund, Luxembourg in the frame of project FNR/A11/02. 
J.M.R.P. acknowledges financial support from grant ENFASIS (FIS2011-22644, Spanish Government). 
This work also benefited from the COST Action MP1209. 


\end{document}